\documentclass[12pt]{article}
\usepackage{amsmath}
\usepackage{latexsym}
\usepackage{amssymb}
\usepackage{a4wide}
\usepackage{bbm}
\usepackage{epsfig}

\newcommand{\E}{\text{e}}

\newcommand{\re}[1]{(\ref{#1})}
\begin{document}
\newcommand{\Nc}{N_{\text{c}}}
\newcommand{\meff}{m_{\text{eff}}}
\newcommand{\Geff}{\Gamma_{\text{eff}}}
\newcommand{\Leff}{{\cal L}_{\text{eff}}}
\newcommand{\rt}{\tilde{\rho}}
\newcommand{\rtc}{\tilde{\rho}_{\text{cr}}}
\newcommand{\rtm}{\tilde{\rho}_{\text{min}}}
\newcommand{\rmin}{\rho_{\text{min}}}
\newcommand{\vf}{\varphi}
\newcommand{\du}{\dot{u}_k}
\newcommand{\duL}{\dot{u}_\Lambda}
\newcommand{\ddu}{\ddot{u}_k}
\newcommand{\lN}{large-$N$ }
\unitlength=1mm
\title{\bf Flow equation for Halpern-Huang directions of scalar O($N$)
  models} 
\author{Holger Gies\\
  \small\it Institut f\"ur theoretische Physik, Universit\"at T\"ubingen,\\
  \small\it 72076 T\"ubingen, Germany}
\maketitle
\begin{abstract}
  A class of asymptotically free scalar theories with O($N$) symmetry,
  defined via the eigenpotentials of the Gaussian fixed point
  (Halpern-Huang directions), are investigated using renormalization
  group flow equations. Explicit solutions for the form of the
  potential in the nonperturbative infrared domain are found in the
  large-$N$ limit. In this limit, potentials without symmetry breaking
  essentially preserve their shape and undergo a mass renormalization
  which is governed only by the renormalization group distance
  parameter; as a consequence, these scalar theories do not have a
  problem of naturalness. Symmetry-breaking potentials are found to be
  ``fine-tuned'' in the large-$N$ limit in the sense that the
  nontrivial minimum vanishes exactly in the limit of vanishing
  infrared cutoff: therefore, the O($N$) symmetry is restored in the
  quantum theory and the potential becomes flat near the origin.
  
\end{abstract}

\section{Introduction}
Common belief holds that only polynomial interactions up to a certain
degree depending on the spacetime dimension are renormalizable, in the
sense that interactions of even higher order require an infinite
number of subtractions in a perturbative analysis. This can be
attributed to the implicit assumption that the higher-order couplings,
which in general are dimensionful, set independent scales. Such
nonrenormalizable theories can only be defined with a cutoff scale
$\Lambda$, while the unknown physics beyond the cutoff is encoded in
the (thereby independent) values of the couplings.

Starting from the viewpoint that the cutoff $\Lambda$ is the only
scale in the theory, Halpern and Huang \cite{halp94a,halp94b} pointed
out the existence of theories with higher-order and even nonpolynomial
interactions within the conventional setting of quantum field theory.
This happens because the higher-order couplings, by assumption, are
proportional to a corresponding power of $1/\Lambda$ and therefore die
out sufficiently fast in the limit $\Lambda\to \infty$; the theories
remain perturbatively renormalizable in the sense that infinitely many
subtractions are not required. Perhaps most important, Halpern and
Huang so discovered nonpolynomial scalar theories which are
asymptotically free, offering an escape route to the ``problem of
triviality'' of standard scalar theories \cite{trivial}.

To be more precise, Halpern and Huang analyzed the renormalization
group (RG) trajectories for the interaction potential in the vicinity
of the Gaussian fixed point. The exact form of the potential was
left open by using a Taylor series expansion in the field as an
ansatz. Employing the Wegner-Houghton \cite{wegn} (sharp-cutoff)
formulation of the Wilsonian RG, the eigenpotentials, i.e., tangential
directions to the RG trajectories at the Gaussian fixed point, were
identified in linear approximation. While the standard polynomial
interactions turn out to be irrelevant as expected, some nonpolynomial
potentials which increase exponentially for strong fields prove to be
relevant perturbations at the fixed point. For the irrelevant
interactions, the Gaussian fixed point is infrared (IR) stable,
whereas the relevant ones approach this fixed point in the ultraviolet
(UV). Possible applications of these new relevant directions are
discussed in \cite{halp94a} for the Higgs model and
in \cite{brac00} for quintessence. Further nonpolynomial potentials
and their applications in Higgs and inflationary models have been
investigated in \cite{lang98}. 

Considering the complete RG flow of such asymptotically free theories
from the UV cutoff $\Lambda$ down to the infrared, the Halpern-Huang
result teaches us only something about the very beginning of the flow
close to the cutoff and thereby close to the Gaussian fixed point.
Each RG step in a coarse-graining sense ``tends to take us out of the
linear region into unknown territory'' \cite{halp94b}. It is the
purpose of the present work to perform a first reconnaissance of this
territory with the aid of the RG flow equations for the ``effective
average action'' \cite{wett93}. In this framework, the standard
effective action $\Gamma$ is considered as the zero-IR-cutoff limit of
the effective average action $\Gamma_k[\phi]$ which is a type of
coarse-grained free energy with a variable infrared cutoff at the mass
scale $k$. $\Gamma_k$ satisfies an exact renormalization group
equation, and interpolates between the classical action
$S=\Gamma_{k\to \Lambda}$ and the standard effective action
$\Gamma=\Gamma_{k\to 0}$. 

In this work, we identify the classical action $S$ given at the cutoff
$\Lambda$ with a scalar O($N$) symmetric theory defined by a standard
kinetic term and a generally nonpolynomial potential of Halpern-Huang
type. Therefore, we have the following scenario in mind: at very high
energy, the system is at the UV stable Gaussian fixed point. As the
energy decreases, the system undergoes an (unspecified) perturbation
which carries it away from the fixed point initially into some
tangential direction to one of all possible RG trajectories. We
assume that this perturbation occurs at some scale $\Lambda$ which
then sets the only dimensionful scale of the system. Any other
(dimensionless) parameter of the system should also be determined at
$\Lambda$; for the Halpern-Huang potentials, there are two additional
parameters: one labels the different RG trajectories; the other
specifies the ``distance'' scale along the trajectory. Finally, the
precise form of the potential at $\Lambda$ serves as the boundary
condition for the RG flow equation which governs the behavior of the
theory at all scales $k\leq \Lambda$. 

Since the RG flow equations for $\Gamma_k$ are equivalent to an
infinite number of coupled differential equations of first order, a
number of approximations (truncations) are necessary to arrive at
explicit solutions. In the present work, we shall determine the RG
trajectory $k\to \Gamma_k$ for $k\in [0,\Lambda]$ explicitly only in
the large-$N$ limit which simplifies the calculations considerably.

The paper is organized as follows: Sec.~\ref{HH}, besides introducing
the notation, briefly rederives the Halpern-Huang result in the
language of the effective average action, generalizing it to
a nonvanishing anomalous dimension. Sec.~\ref{largeN} investigates
the RG flow equation for the Halpern-Huang potentials in the large-$N$
limit, concentrating on $d=3$ and $d=4$ spacetime dimensions; here, we
emphasize the differences to ordinary $\phi^4$ theory particularly in
regard to mass renormalization and symmetry-breaking properties.
Sec.~\ref{conclusions} summarizes our conclusions and discusses open
questions related to finite values of $N$. 


As an important caveat, it should be mentioned that the results of
Halpern and Huang have been questioned (see \cite{question} and also
\cite{bagn00}), and these questions raised also affect the present
work. To be honest, we have hidden the problems in the ``scenario''
described above in which an ``unspecified'' perturbation controls the
shift of the system from the Gaussian fixed point (the continuum
limit) to the cutoff scale $\Lambda$ along a {\em tangential}
direction. But since the cutoff scale $\Lambda$, though large, is not
at all infinitesimally separated from the Gaussian fixed point, this
tangential approximation is probably not sufficient to stay on the
true renormalized trajectory during the shift. Not only the tangent
but also all (infinitely many) curvature moments of the trajectory had
to be known in order to find an initial point right on the
renormalized trajectory at $\Lambda$. This point would correspond to a
so-called ``perfect action'' \cite{Hasenfratz:1998bb}. Of course, this
requires an infinite number of conditions to be imposed on the initial
action at $\Lambda$ which we cannot specify. In conventional field
theories, this problem is solved by adjusting (fine-tuning) the
initial action close to the unstable Gaussian fixed point, leaving
open only one a priori chosen relevant direction to the flow. But in
the present case, there is an infinite number of relevant directions
corresponding to the continuum of possible Halpern-Huang directions,
and thus it seems impossible to single out only one relevant direction
while frustrating the others by tuning infinitely many parameters.  In
other words, upon studying the flow from $\Lambda$ down to zero within
our scenario, the continuum limit of our system remains unspecified,
and therefore one important ingredient to a complete field theoretic
system is missing.
 
With these reservations in mind, we nevertheless believe that there
are some lessons to be learned from the application of the RG flow
equations to such potentials.

\section{Scalar O($N$) theories close to the Gaussian fixed point}
\label{HH}

Concerning the investigation of the RG flow equation for the Euclidean
effective average action in $d$ dimensions, we closely follow the
original work of Wetterich \cite{wett93}. Polynomial potentials and
the large-$N$ limit to be discussed later have been explored in
\cite{tetr94} and \cite{Tetradis:1996br} in the effective average
action approach. A comprehensive review and an extensive list of
references on this subject can be found in \cite{berg00}. The
effective average action can be expanded in terms of all possible
O($N$) invariants,
\begin{equation}
\Gamma_k[\phi] =\int d^dx\left\{ U_k(\rho) + \frac{1}{2} Z_k(\rho)\,
  \partial_\mu \phi^b \partial_\mu \phi^b + \frac{1}{4} Y_k(\rho)\,
  \partial_\mu \rho \partial_\mu \rho+\dots\right\}, \label{1}
\end{equation}
where $\rho:=\frac{1}{2} \phi^b \phi^b$, $b=1\dots N$ labels the real
components of the scalar field, and the dots represent terms involving
higher derivatives; for convenience, we shall always assume that $d>2$
during the calculation. Halpern and Huang derived their result in the
``local-potential approximation'' which is constituted by setting the
wave function renormalization constant $Z_k\equiv 1$ and neglecting
$Y_k$ and higher-derivative terms. In the present work, we shall
generalize their result to a $k$-dependent $Z_k$ which is parametrized
by the anomalous dimension,
\begin{equation}
\eta:= -\partial_t \ln Z_k, \quad\text{where}\quad \partial_t\equiv
k\frac{d}{dk} \label{2}
\end{equation}
denotes the derivative with respect to the RG ``time'', $t\in
]-\infty, 0]$\footnote{Note that our convention follows \cite{wett93}
  and thus is opposite to the one used by Halpern and Huang in
  \cite{halp94a,halp94b} and also by the author of \cite{peri95}:
  $t=-t_{\text{HH}}$.} $t=\ln k/\Lambda$.  Here we neglect $Y_k$ and
any $\rho$ dependence of $Z_k$.  Following \cite{wett93}, the RG flow
equation for the effective average potential $U_k(\rho)$ can be
written as
\begin{equation}
\partial_t U_k(\rho) =\frac{1}{2} \int \frac{d^d q}{(2\pi)^d} \,
\partial_t R_k\, \left( \frac{N-1}{Z_k q^2 +R_k +U_k'(\rho)} +
  \frac{1}{Z_k q^2 +R_k +U_k' +2\rho U_k''(\rho)} \right),
\label{3}
\end{equation}
where the prime denotes the derivative with respect to the argument
$\rho$. The cutoff function $R_k=R_k(q^2)$ is to some extent an
arbitrary positive function that interpolates between $R_k(q^2) \to
Z_k k^2$ for $q^2\to 0$ and $R_k(q^2)\to 0$ for $q^2\to \infty$. It
suppresses the small-momentum modes by a mass term $k^2$ acting as the
IR cutoff. In Eq.~\re{3} the distinction between the $N-1$ ``Goldstone
modes'' and the ``radial mode'' is visible.

Provided that $\eta$ is given (which we shall always assume in the
this work), the flow of the effective potential $U_k$
(and thus of the effective action $\Gamma_k$ in the present
approximation) is determined by Eq.~\re{3}. Even if $\eta$ is
neglected, Eq.~\re{3} produces qualitatively good results for polynomial
effective potentials in $d>2$ \cite{berg00}. We expect similar
behavior for nonpolynomial potentials. 

The Halpern-Huang result can be rederived by assuming that the system
is close to the Gaussian fixed point so that the effective potential
and its derivatives are small. Linearizing the right-hand side of
Eq.~\re{3} with respect to the potential and its derivatives gives
\begin{equation}
\partial_t U_k(\rho) =-v_d\, \bigl( N U_k'(\rho) +2\rho U_k''(\rho)
\bigr) \int\limits_0^\infty dw\, w^{d/2 -1}\, \frac{\partial_t
  R_k(w)}{(Z_k w+R_k(w))^2} +{\cal O}(U_k^2), \label{4}
\end{equation}
where we introduced the abbreviation $v_d=2^{-(d+1)} \pi^{-d/2}
\Gamma^{-1}(d/2)$, which is related to the volume of $d$ spheres.
It is convenient to remove the explicit $Z_k$ and $k$
dependence by using dimensionless scaling variables:
\begin{eqnarray}
\vf=Z_k^{1/2} k^{1-d/2}\phi,\quad \rt=\frac{1}{2}\vf^2 =Z_k k^{2-d}
\rho, \quad u_k(\vf)=k^{-d}\, U_k(\phi). \label{5}
\end{eqnarray}
In the same spirit, we write for the cutoff function
\begin{equation}
R_k(q^2)=Z_k k^2\, C(q^2/k^2), \label{6}
\end{equation}
where $C(w)$ is a dimensionless function of a dimensionless argument,
satisfying $C(w\to 0) \to 1$ and $C(w\to\infty)\to 0$. Rewriting
Eq.~\re{4} in terms of these variables and taking the RG time
derivative $\partial_t$ on the left-hand side at fixed $\rt$, we
obtain the differential equation 
\begin{equation}
\partial_t u_k(\rt) =-d u_k(\rt) +(d-2+\eta)\rt \du(\rt) -\frac{1}{2}
\kappa \bigl( 2\rt \ddu(\rt) +N \du(\rt)\bigr), \label{7}
\end{equation}
where the dot denotes a derivative with respect to the argument $\rt$,
and the complete cutoff dependence is contained in 
\begin{equation}
\kappa=\kappa(d,\eta;C)=2v_d \int\limits_0^\infty dw \left[ (d-2)
  \frac{w^{d/2-2} C(w)}{w+C(w)} -\eta \frac{w^{d/2-1}
  C(w)}{(w+C(w))^2} \right]. \label{8}
\end{equation}
We are looking for eigenpotentials, i.e., tangential directions
to the RG flow of the scaling form $u_k\sim \E^{-\lambda t}$, where
$\lambda$ classifies the possible directions and distinguishes between
irrelevant ($\lambda<0$), marginal ($\lambda=0$) and relevant
($\lambda>0$) perturbations away from the Gaussian fixed
point. Solutions of this form can be given in terms of the Kummer
function $M$ \cite{abra}
\begin{equation}
u_k(\rt)=-\E^{-\lambda t}\, \frac{2 \kappa r}{d-\lambda} \left[
  M\left(\frac{\lambda -d}{d-2+\eta}, \frac{N}{2};
  \frac{d-2+\eta}{\kappa} \,\rt\right)-1\right]. \label{9}
\end{equation}
For given dimension, $N$, cutoff specification and anomalous
dimension, the Halpern-Huang potential \re{9} depends on two
dimensionless parameters: $\lambda$ and $r$. The latter sets a
``distance'' scale along the RG trajectories; since it is an overall
factor, the position of possible extrema of $u_k$ are independent of
$r$. 

To make contact with the literature, we note that we rediscover the
results of Periwal \cite{peri95} in the limit $\eta=0$, where the
Halpern-Huang result was generalized to arbitrary cutoffs within the
Polchinski RG approach \cite{polc84}. The results of Halpern and Huang
are recovered by employing a sharp cutoff, for which $\kappa$ is
related to the volume of the $d-1$ dimensional sphere\footnote{The
  sharp cutoff limit of Eq.~\re{8} has to be defined carefully;
  details can be found in \cite{berg00,bagn00}.}:
\begin{equation}
\kappa(d,\eta=0,C_{\text{sc}})=\frac{\text{vol.}(S^{d-1})}{(2\pi)^d}.
\label{11}
\end{equation}

Various representations for the Kummer function
$M(\alpha,\beta;x)$ exist in the literature \cite{abra}; for further
discussion, it is useful to replace the parameter $\lambda$ by the
combination 
\begin{equation}
a:=1+ \frac{\lambda -d}{d-2+\eta}. \label{12}
\end{equation}
Then, Eq.~\re{9} reduces to standard polynomial potentials of degree
$n$ in $\rt$ ($2n$ in $\vf$) if $a=-n+1$; for all such polynomial
potentials, the Gaussian fixed point is IR stable. For $a=1$, the
potential vanishes, and for any other value of $a$, the potential is
nonpolynomial. For these cases, the asymptotic behavior for large
third argument $x$ is given by an exponential increase
\begin{equation}
M(\alpha,\beta;x)\simeq \frac{\Gamma(\beta)}{\Gamma(\alpha)}\,
x^{(\alpha-\beta)}\, \E^{x} \bigl( 1+{\cal
  O}(x^{-1})\bigr). \label{10}
\end{equation} 
The Gaussian fixed point is UV stable ($\lambda>0$) for
\begin{equation}
a>-\frac{2-\eta}{d-2+\eta}, \label{13}
\end{equation}
(as long as $d-2+\eta>0$). A particularly interesting case is given by
the parameter set
\begin{equation}
-1<a<0, \quad r<0, \label{14}
\end{equation}
for which the eigenpotential Eq.~\re{9} is nonpolynomial and develops
a minimum, inducing spontaneous symmetry breaking.

To conclude our derivation of the Halpern-Huang results, we mention
that in the particular case of $N=1$ there exist (physically
admissible) solutions to Eq.~\re{7} which are odd under $\vf\to-\vf$
\cite{brac00}. The linearized flow equation has also been studied from
a different perspective employing its similarity to a Fokker-Planck
form \cite{bonn00}.

According to the scenario outlined in the introduction, we shall now
consider the potentials found in Eq.~\re{9} taken at $t=0$
($k=\Lambda$) as the boundary condition for the complete flow equation
\re{3}. Provided that the anomalous dimension $\eta$ is only weakly
dependent on $k$ and bounded (as is the case, e.g., for polynomial
interactions in $d>2$), some features can immediately be read off from
Eq.~\re{3}: for nonpolynomial potentials with exponential asymptotics
given by Eq.~\re{10}, the denominators on the right-hand side of the
flow equation \re{3} vanish exponentially for large values of $\rho$.
Therefore, $\partial_t U_k(\rho)\to 0$ for large $\rho$, and the flow
halts, leaving $U_k$ essentially unchanged.

In particular, for symmetry-preserving potentials with a minimum at
$\rho=0$ and $a>1$, we may expect a rather unspectacular flow: for
large $\rho$, the above argument holds, whereas for small $\rho$, we
may always find a small region where the linearization of the flow
equation is a good approximation; there, the Halpern-Huang potential
will still be an appropriate approximation. Therefore, these
potentials are expected to behave stiffly under the flow.

For potentials with a minimum at nonvanishing $\rho$ (spontaneous
symmetry breaking) with $-1<a<0$, the asymptotics for large $\rho$ 
will also stop the flow. However, the flow of $U_k$ near the
nontrivial minimum can be more complicated, since $U_k'$ and $U_k''$
are no longer monotonic functions in this region. To the right of the
minimum, these potentials may also be stiff under the flow, but the
region around the origin and the minimum appear as a loose end.

These heuristic arguments will be worked out and
confirmed in the following section in the large-$N$ limit. 

\section{RG flow of Halpern-Huang theories in the large-$N$ limit}
\label{largeN}

For solving flow equations for the effective average potential of the
type of Eq.~\re{3}, several techniques have been developed. Of course,
it is always possible to search numerically for solutions by putting
the differential equation on a computer; in fact, if one is looking
for accurate results, this is the most appropriate option. However,
since the potentials under consideration exhibit an exponential
increase, straightforward numerics may come to its limits and a clever
variable substitution has to be guessed. 

Another possibility is to expand the potential in terms of a complete
set of functions and decompose the flow equation into differential
equations for the $k$-dependent coefficients (generalized
couplings). Here, a choice for a useful set of functions again has to
be guessed; obviously, the polynomials as the standard choice are of
no use, because the important information is contained in the
nonpolynomial nature of the potential. 

Therefore, we decide to work in the large-$N$ limit which puts
no a priori restrictions on the form of the potential and allows for
a complete integration of the flow equation. Of course, the validity
of the results for finite values of $N$ can hardly be controlled at
this early stage.

\subsection{RG flow equation in the \lN limit}


In the \lN limit, the RG flow equation \re{3} for the potential
simplifies considerably; here we shall follow the presentation given
in \cite{Tetradis:1996br} and \cite{berg00}. Not only does the
anomalous dimension $\eta$ vanish \cite{zinn89}, but so does the
influence of higher derivative terms ($Y_k,\dots$). Moreover, the
Goldstone modes dominate the right-hand side of Eq.~\re{3} and any
contribution from the radial mode can be neglected (this essentially
changes the order of the differential equation).

For technical reasons, one finally chooses a sharp cutoff function
$R_k$ and decides to consider the flow equation for the {\em
  derivative} of the potential. In dimensionless variables, the \lN
limit of the flow equation reads
\begin{equation}
2\du=-\frac{\partial \du}{\partial t} + \left( (d-2) \rt -\frac{2v_d
    N}{1+\du} \right) \frac{\partial \du}{\partial \rt}. \label{15}
\end{equation}
Of course, this equation can be obtained directly from the
sharp-cutoff formulation of the RG and has already been studied by
Wegner and Houghton \cite{wegn}; further investigations of the
Wegner-Houghton approach have been made in \cite{Comellas1997}.

Following \cite{Tetradis:1996br}, this partial differential equation
of first order can be solved using the standard method of
characteristics and we find that the solution $\du(\rt)$ has to
satisfy the equation
\begin{equation}
\rt =s(\du)\, \E^{-(d-2) t} -v_d N\, I(d,t;\du), \label{16}
\end{equation}
where $I(d,t;\du)$ is defined by the integral
\begin{equation}
I(d,t;\du):= \E^{-(d-2)t} \int\limits_0^{\exp(-2t)} dw\,
\frac{w^{-d/2}}{1+\E^{2t} \, \du\, w}. \label{17}
\end{equation}
This function is studied in App. A and explicit representations for
$d=3$ and $d=4$ are given. The function $s(\du)$ is implicitly defined
by the equation
\begin{equation}
\dot{u}_\Lambda(s)=\du\E^{2t}\quad \Longrightarrow\quad s\equiv
s(\du), \label{18} 
\end{equation}
where $\dot{u}_\Lambda(s)$ represents the boundary condition for the
flow equation at $k=\Lambda$ ($t=0$); here, $s$ as a variable
parametrizes the boundary condition and corresponds to the $\rt$ axis
at $t=0$ in the $\rt,t$ plane. It is exactly Eq.~\re{18} that is to be
inserted into Eq.~\re{16}, where the nonpolynomial potentials enter
the investigation.

Now the route to an explicit solution is clear: (i) we specify the
boundary condition via Eq.~\re{18}, (ii) insert this and an explicit
representation for $I(d,t;\du)$ into Eq.~\re{16}, and ``solve'' (or
invert) the resulting equation for $\du(\rt)$. However, in practise,
some complications are encountered: e.g., inverse functions of such
complicated objects as the Kummer function $M$ are not easily
obtainable. But the \lN limit comes to the rescue once more as
demonstrated in the next subsection.

Let us finally extract the flow of a possible minimum of the
potential which is defined by $\du(\rtm)=0$; from Eq.~\re{16}, we can
easily extract that
\begin{equation}
\rtm(k)=\rtm(\Lambda)\,\E^{-(d-2)t} -v_d N\,I(d,t;0), \label{19}
\end{equation}
where $\rtm(\Lambda)$ denotes the minimum of the potential at
$k=\Lambda$ (t=0), i.e., the minimum of the Halpern-Huang
potential (in the \lN limit); by construction, it is identical to
$\rtm(\Lambda)=s(\du(\rtm)=0)$. The function $I(d,t;0)$ can be read
off from Eqs. \re{A3} and \re{A4} of the appendix ($I(d,t;0)\equiv
i_0(d,t)$). Reinstating dimensionful quantities (cf. Eq.~\re{5}), we
find for the flow of a minimum of the potential
\begin{equation}
\rmin(k)=\rmin(\Lambda) -\rtc \bigl(\Lambda^{d-2} -k^{d-2} \bigr),
\quad \rtc:= \frac{2v_d N}{d-2}. \label{20}
\end{equation}
Here, we introduced a ``critical'' (dimensionless) field strength
$\rtc$. Of course, Eq.~\re{20} is well known in the literature
\cite{berg00} and makes no particular reference to the type of
potential under consideration. The only place where the potential type
enters is the position of the initial minimum $\rmin(\Lambda)$. If
$\rmin(\Lambda)>\rtc\Lambda^{d-2}$, then the classical as well as the
quantum theory exhibit spontaneous symmetry breaking, since
$\rmin(k\to 0)>0$; if $\rmin(\Lambda)<\rtc\Lambda^{d-2}$ the quantum
theory will preserve O($N$) symmetry. Finally, if
$\rmin(\Lambda)=\rtc\Lambda^{d-2}$ the classical potential $U_\Lambda$
is ``fine-tuned'' in such a way that the theory shows symmetry
breaking for finite values of $k$, but restores O($N$) symmetry in the
limit $k\to 0$; additionally, the potential has a vanishing mass term:
$M^2:=U_{k\to 0}'(0)=0$ (by construction)\footnote{In the language of
  statistical mechanics, the theory is exactly at the critical
  temperature.}.  

In standard $\phi^4$ theory, the position of the minimum
$\rmin(\Lambda)$ of $U_\Lambda$ can be chosen at will by an
appropriate tuning of the negative mass term and the coupling. By
contrast, for Halpern-Huang potentials, once the precise type of the
potential is chosen by fixing $\lambda$ (or $a$), there is no
parameter left for any fine-tuning, since a possible minimum (for
theories with $-1<a<0$) is independent of the last free parameter $r$
in Eq.~\re{9}. The question as to whether a symmetry-breaking quantum
theory of the Halpern-Huang potentials exists has to be answered
by determining the position of the initial minimum
$\rmin(\Lambda)$. This will also be investigated in the next
subsection in the \lN limit.

\subsection{Large-$N$ limit of the Halpern-Huang potentials}
\label{lNHH}

In our scenario, the Halpern-Huang potential enters the flow equation
as its boundary condition at the cutoff. Since the flow equation is
considered in the \lN limit, it is not only useful to insert the \lN
limit of the Halpern-Huang potential into Eq.~\re{16}, but it is
mandatory for reasons of consistency. Otherwise, nonleading \lN
information would be mixed with \lN behavior, introducing some
arbitrariness into this approximation. 

From Eq.~\re{9}, we read off that the parameter $N$ occurs only in the
second argument of the Kummer function. Unfortunately, we could not
find any asymptotic expression for the Kummer function with large
second argument in the literature. Instead of investigating this limit
in terms of some appropriate series or integral representation, which
might involve awkward interchanges of limiting processes, we shall use
a more physically motivated approach: since the Kummer function was
identified with the tangential direction to the RG flow in the
vicinity of the Gaussian fixed point, its \lN approximation should
naturally be deducible from a \lN study of the same subject. In other
words, the desired function has to be a solution to the linearized
flow equation in the \lN limit.

For technical reasons, we again turn to the derivative of the
potential with respect to $\rt$. Then, the desired differential
equation is obtained by linearizing Eq.~\re{15}. Looking for
potentials which satisfy the eigenpotential scaling condition
$-\partial_t \du=\lambda\, \du$, the \lN Halpern-Huang equation reads
\begin{equation}
\frac{d \dot{u}_\Lambda(\rt)}{d\rt} =-\frac{a}{\rt-\rtc}\,
\dot{u}_\Lambda(\rt), \label{21}
\end{equation}
where we again traded $\lambda$ for the parameter $a$ as defined in
Eq.~\re{12}. Additionally, we made use of the ``critical'' field strength
$\rtc$ defined in Eq.~\re{20}.
 
Equation \re{21} can easily be solved for the various boundary
conditions; let us begin with a symmetry-preserving potential
satisfying $\duL(\rt)>0$ and $a>0$:
\begin{equation}
\duL(\rt)=\upsilon_+\, \left( \frac{\rtc}{\rtc -\rt} \right)^a, \quad
\upsilon_+:= \duL(0), \label{22}
\end{equation}
where the initial value $\upsilon_+$ also satisfies
$\upsilon_+>0$. The parameter $\upsilon_+$ is, of course, the \lN
analogue of the distance parameter $r$ in Eq.~\re{9}, $r\to(N/4)
\upsilon_+$.  

\begin{figure}
\begin{picture}(145,52) 
\put(0,0){
\epsfig{figure=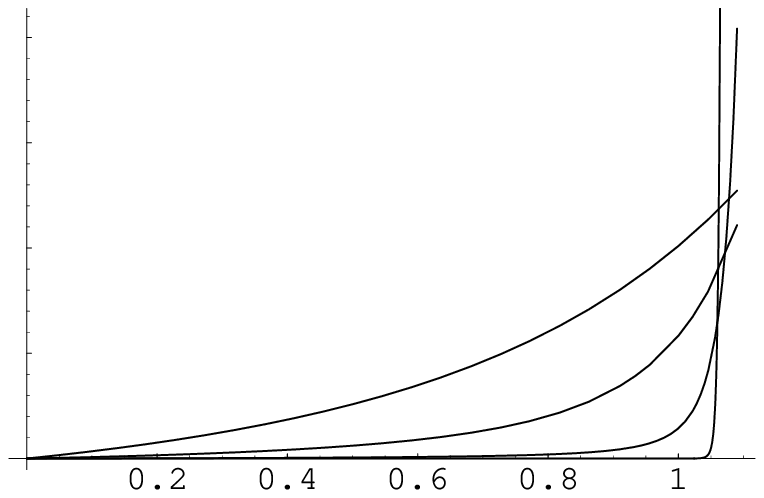,width=7.7cm}}
\put(0,52){(a)}   
\put(49,18){$N$=10}
\put(61,13){100}
\put(77,40){1000}
\put(64,45){10000}
\put(7,49){$\du(\rt)$}
\put(80,3){$\frac{\rt}{\rtc}$}
\put(82,0){
\epsfig{figure=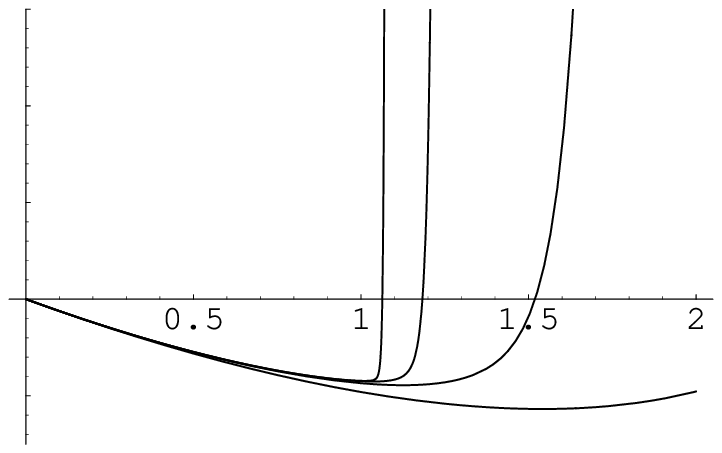,width=7.7cm}}
\put(82,52){(b)}
\put(150,7){10}
\put(148,40){100}
\put(134,40){1000}
\put(110,40){$N$=10000}
\put(92,46){$\du(\rt)$}
\put(156,20){$\frac{\rt}{\rtc}$}
\end{picture}
\caption{Large-$N$ behavior of (a) a symmetry-preserving ($a=2$) and (b) a
  symmetry-breaking Halpern-Huang potential ($a=-1/2$) in $d=4$
  dimensions versus $\rt/\rtc$. For $N\to \infty$, the potentials
  develop a ``wall'' at $\rt=\rtc$. In (a), the normalization of the
  potentials is appropriately adapted for illustrative purposes.}
\label{Fig1} 
\end{figure}

At first sight, one may doubt the validity of Eq.~\re{22} as the \lN
limit of Eq.~\re{9}, because it diverges at the critical field
strength $\rt\to\rtc$. Nevertheless, this indeed reflects the behavior
of the Kummer function for large second argument, as can be checked
numerically (see Fig. \ref{Fig1}(a)) \cite{math}. The critical field
strength $\rtc$ marks the point where the asymptotic exponential
increase (cf. Eq.~\re{10}) sets in; and for larger values of $N$, the
slope increases without bound\footnote{This might be the reason why
  there is no \lN limit of the Kummer function in the literature: it
  cannot be defined for arbitrary values of $\rt$.}. Of course, a
potential that diverges for finite values of its argument is usually
considered as inadmissible in field theory (see, e.g., \cite{wegn} and
\cite{Hasenfratz:1986dm}); however, in the present
case, we take the viewpoint that this potential wall at $\rt=\rtc$
only symbolizes the exponential increase of the potential for finite
values of $N$. 

Let us now turn to the solution of Eq.~\re{21} for the
symmetry-breaking potentials with ($-1<a<0$). In the inner region of
the potential where $\duL<0$, we obtain the solution
\begin{equation}
\duL(\rt)=\upsilon_- \left( \frac{\rtc -\rt}{\rtc} \right)^{-a}, \quad
\upsilon_-:= \duL(0), \quad |a|=-a, \label{23}
\end{equation}
where the initial value this time satisfies $\upsilon_-<0$. The
derivative of the \lN Halpern-Huang potential has a zero at
$\rt=\rtc$, so that the potential itself exhibits a minimum at this
position. Now let us turn to the \lN limit of the symmetry-breaking
potential to the right of the minimum $\rt>\rtc$ where $\duL>0$. As a
matter of fact, the unique solution of Eq.~\re{21} increases only very
slowly, $\duL\sim \rt^{|a|}$ for $\rt \to\infty $ and $-1<a<0$. This
does certainly not reflect the expected exponential increase; hence
this solution has to be discarded. Even if a more appropriate solution
existed for $\rt>\rtc$, we would not be able to match them properly at
$\rt=\rtc$, because the second derivative of the potential diverges at
this point. In view of the results for the symmetry-preserving
potential and owing to the fact that the asymptotic exponential
increase for both types of potentials is the same (see Eq.~\re{10}),
the only possibility for the \lN limit is to continue the potential at
$\rt=\rtc$ by a potential wall at $\rt=\rtc+0^+$. Again, a numerical
analysis of the full Kummer function for large $N$ confirms this
conjecture, as is depicted in Fig. \ref{Fig1}(b).

This concludes our \lN analysis of the Halpern-Huang potential; note
that both types of the potential are formally equivalent, so that we
combine them in the notation
\begin{equation}
\duL(\rt)= \upsilon \left( \frac{\rtc}{\rtc-\rt} \right)^a, \quad
\upsilon:= \duL(0), \label{24}
\end{equation}
where $\upsilon$ stands for $\upsilon_+$ or $\upsilon_-$ and $a\in
\mathbbm{R}$. Actually, this also covers a $\phi^4$ potential for
$a=-1$. 

\subsection{Nonperturbative evolution of the Halpern-Huang potentials}
Now we are in a position to study the flow of the Halpern-Huang
potentials from $k=\Lambda$ into the infrared regime $k\to 0$ in the
\lN limit. The missing piece of information to be inserted into
Eq.~\re{16} is given by the inverse of Eq.~\re{24}:
\begin{equation}
\duL(s)= \upsilon \left( \frac{\rtc}{\rtc-s} \right)^a, \quad
\stackrel{\re{18}}{\Rightarrow}\quad 
s(\du)=\rtc\left[1-\left(\frac{\upsilon}{\du \E^{2t}} \right)^{1/a}
\right]. \label{25}
\end{equation}
Employing the representation \re{A3} for the function $I(d,t;\du)$, we
find that the derivative of the potential has to satisfy the equation
\begin{equation}
0=(\rtc-\rt) -\rtc \left(\frac{\upsilon}{\du(\rt)}\right)^{1/a}
\E^{-(\lambda/a) t} +\frac{d-2}{2} \rtc\,\du(\rt)\, J(d,t;\du),
\label{26}
\end{equation}
where the function $J(d,t;\du)$ is defined in Eq.~\re{A5}. Finally,
Eq.~\re{26} has to be solved for $\du(\rt)$, which we shall do in the
limit $k\to0$ ($t\to-\infty$) for various cases in order to obtain the
complete quantum effective potential. The following consideration will
serve as a guide to the necessary approximations: at the cutoff
$k=\Lambda$, the dimensionful potential is of the order of the cutoff
$U_\Lambda'\sim \Lambda^2$. For small deviations from the cutoff,
$k/\Lambda\lesssim 1$, the potential scales according to the
linearized flow equation (Halpern-Huang equation):
$\du\sim\E^{-\lambda t}$. Then, the dimensionful potential scales as
$U_k'\sim \E^{-(\lambda-2) t}\sim \E^{-(d-2)a t}$. Therefore, if $a>0$
(symmetry-preserving potentials), $U_k'$ increases as we approach the
infrared, whereas if $a<0$ (symmetry-breaking potentials), $U_k'$
decreases towards the infrared. Of course, this argument holds
strictly close to $k\simeq \Lambda$ only, but it turns out to
reproduce the unique consistent approximation schemes for extracting
analytical results. 

\subsubsection{Symmetry-preserving potentials in $d=4$}

Let us first consider the $d=4$ potentials with $a>0$ and
$U_\Lambda'>0$ that exhibit no symmetry breaking at the
cutoff. Employing Eq.~\re{A6} and reinstating dimensionful quantities
via Eq.~\re{5}, Eq.~\re{26} reads, after neglecting terms of order
$k^2/U_k'$ in the limit $k\to 0$ ($U_{k\to 0}'\equiv U'$):
\begin{equation}
0=\rtc\, U'(\rho) \ln \left(1+\frac{\Lambda^2}{U'(\rho)}\right)
-\rho -\rtc \Lambda^2 \left(\frac{U_\Lambda'(0)}{U'(\rho)}
\right)^{1/a} , \label{27}
\end{equation}
where $U_\Lambda'(0)=\upsilon_+\, \Lambda^2\equiv M^2_\Lambda$ denotes
the mass of the theory at the cutoff. Let us study Eq.~\re{27} in two
limits: first, at $\rho$ close to $\rtc \Lambda^2$, and secondly at
the origin $\rho\to 0$. 

At $\rho$ close to the potential wall at $\rtc \Lambda^2$, the
potential diverges, and we can approximate $\Lambda^2\ll U'(\rho\to
\rtc\Lambda^2)$, leading us to
\begin{equation}
U'(\rho\to \rtc\Lambda^2) =U_\Lambda'(0) \left(
  \frac{\rtc\Lambda^2}{\rtc\Lambda^2 -\rho}\right)^a. \label{28}
\end{equation}
In this limit, the effective potential $U\equiv U_{k\to 0}$ remains
formally identical to the \lN Halpern-Huang potential
(cf. Eq.~\re{22})! This confirms our heuristic argument that the
potential behaves stiffly under the flow in the region where it
increases exponentially. 

\begin{figure}
\begin{picture}(145,52) 
\put(0,0){
\epsfig{figure=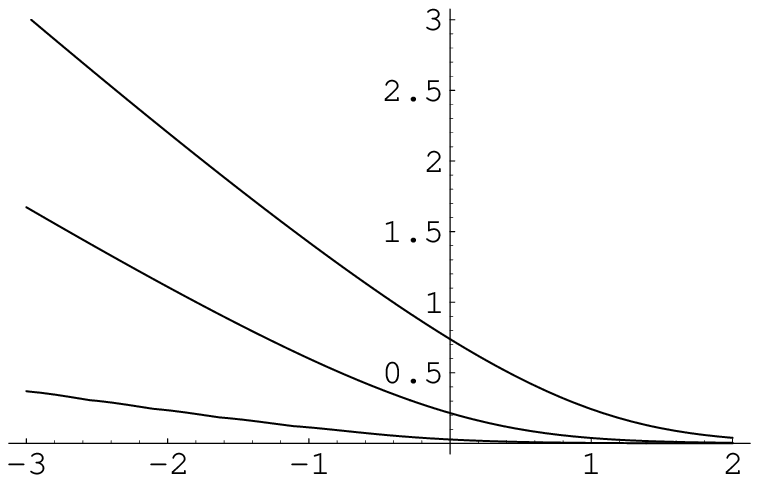,width=7.5cm}}
\put(0,52){(a): $d=4$}   
\put(35,50){$\log_{10} \frac{M^2}{M_\Lambda^2}$}
\put(61,-2){$\log_{10} \upsilon_+$}
\put(2,10){$a=0.2$}
\put(5,27){$a=2$}
\put(12,40){$a=20$}
\put(84,0){
\epsfig{figure=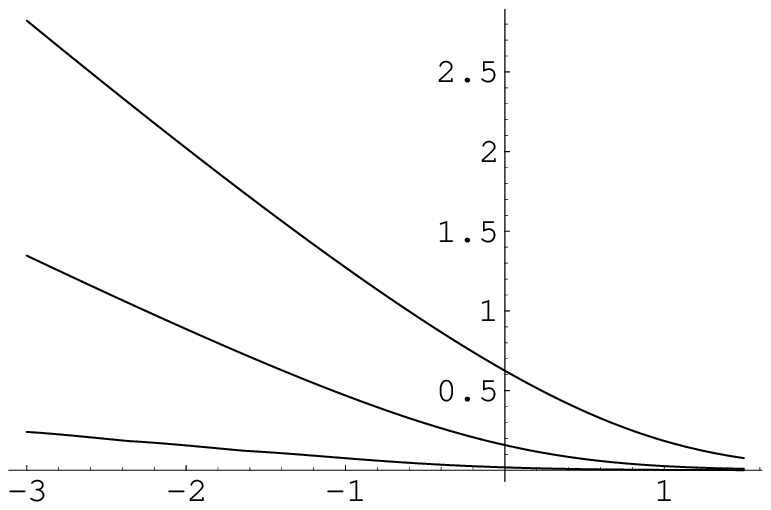,width=7.5cm}}
\put(84,52){(b): $d=3$}
\put(147,-2){$\log_{10} \upsilon_+$}
\put(127,50){$\log_{10} \frac{M^2}{M_\Lambda^2}$}
\put(86,9){$a=0.2$}
\put(87,25){$a=2$}
\put(101,38){$a=20$}  
\end{picture}
\caption{Double-logarithmic plot of the renormalized-to-cutoff mass
  ratio depending on the RG distance parameter $\upsilon_+$ for (a)
  $d=4$ (cf. Eq. \re{29}) and (b) $d=3$ (cf. Eq. \re{33}) for various
  values of $a$.}
\label{Fig3}
\end{figure}

Concerning the opposite limit $\rho\to 0$, there would be no mass
renormalization at all, if Eq.~\re{28} were also correct in this
limit, $M^2:=U'(0)\stackrel{?}{=}U_\Lambda'(0)=M^2_\Lambda$. However,
in this limit, the approximation $\Lambda^2\ll U'$ no longer holds,
and instead we deduce from Eq.~\re{27} the transcendental equation
\begin{equation}
1=\upsilon_+\left(\frac{M^2}{M_\Lambda^2} \right)^{(a+1)/a} \ln \left(
  1+ \frac{1}{\upsilon_+} \frac{M_\Lambda^2}{M^2} \right). \label{29}
\end{equation}
Therefore, the mass renormalization is governed by the only free
parameter of the theory, $\upsilon_+$: for large $\upsilon_+$, there
is effectively no renormalization, whereas the renormalized mass $M^2$
exceeds the ``classical'' mass $M_\Lambda^2$ for $\upsilon_+\lesssim
1$. Typical values are $M^2\simeq 10 M_\Lambda^2$ for
$\upsilon_+=0.01$ and $a=2$; for larger values of the RG trajectory
parameter $a$, the mass shift even increases: $M^2\simeq 100
M_\Lambda^2$ for $\upsilon_+=0.01$ and $a=20$. The $M^2/M_\Lambda^2$
relation is plotted against $\upsilon_+$ for various $a$ in Fig.
\ref{Fig3}(a).

By reintroducing the cutoff again via $M_\Lambda^2=v_+\Lambda^2$,
Eq. \re{29} can be interpreted differently by writing
\begin{equation}
v_+= \left( \frac{M^2}{\Lambda^2} \right)^{a+1} \left[ \ln \left(1+
    \frac{\Lambda^2}{M^2} \right) \right]^a. \label{30}
\end{equation}
This equation tells us that the physical mass of the theory in the
infrared can easily be much smaller than the cutoff by tenth of orders
of magnitude, provided that $\upsilon_+$ is correspondingly small.
Since $\upsilon_+$ sets the distance scale on the RG trajectory, the
demand for a small value of $\upsilon_+$ is consistent with our
scenario: if we leave the Gaussian fixed point with a very tiny
perturbation $\sim \upsilon_+$ at the high-energy scale $\Lambda$, it
is only {\em natural} to arrive at a low-energy theory with a
similarly tiny mass compared to the cutoff. Moreover, consistency of
our scenario requires $\upsilon_+$ to be small in order to justify the
linearization of the flow equation in deriving the Halpern-Huang
result.

To summarize, the symmetry-preserving Halpern-Huang potential
qualitative does not change its form during the flow into the
infrared; in particular, no symmetry breaking occurs. Only the slope
of the potential at the origin of the theory increases for $k\to0$,
which corresponds to a mass renormalization.

\subsubsection{Symmetry-breaking potentials in $d=4$}
Let us begin with a dimension-independent statement referring to the
position of the minimum of symmetry-breaking Halpern-Huang potentials
with $-1<a<0$: in Subsec.~\ref{lNHH} we learned that the position of
the minimum of the Halpern-Huang potentials in the large-$N$ limit is
independent of the parameters $a$ and $v_-$:
$\rt_{\text{min}}(\Lambda)=\rtc$, or in dimensionful quantities:
$\rmin(\Lambda)= \rtc \Lambda^{d-2}$. According to the discussion
following Eq.~\re{20}, the Halpern-Huang potentials are ``fine-tuned''
in the sense that the minimum vanishes exactly in the infrared limit
$k\to 0$:
\begin{equation}
\rmin(k)=\rtc k^{d-2}. \label{31}
\end{equation}
Therefore, there is no symmetry breaking in the full quantum theory of
Halpern-Huang potentials in the large-$N$ limit. Moreover, since
$M^2=U'(\rho=0)\equiv U_{k=0}'(\rho=0)=0$, the potential is flat at
the origin and the renormalized quantum theory is massless. 

Following the line of argument given below Eq. \re{26}, the inner
region of the potential where $U_k'<0$ decreases towards the infrared;
hence we approximate $|U_k'|/\Lambda^2\ll 1$ in Eq. \re{26} and obtain
the transcendental equation in $d=4$:
\begin{equation}
-U_k'=k^2 -\Lambda^2 \exp \left( -\frac{\rtc k^2-\rho}{\rtc (-U_k')}
 \right). \label{32}
\end{equation}
Here we can read off that $|U_k'|$ is always smaller than $k^2$. This
reflects the approach to convexity of the inner part of the effective
potential. 

To summarize, we have found, on the one hand, that the originally
nontrivial minimum of the potential moves to the origin during the
flow; the inner region of the potential shrinks to a point. On the
other hand, we know from the preceding subsection that the potential
wall at $\rho=\rtc\Lambda^2 +0^+$ does not change its position under
the flow. It remains to be investigated what happens in between the
minimum and the potential wall. Unfortunately, we cannot answer this
question by the large-$N$ version of the flow equation, because we do
not have a boundary condition for this region. At the cutoff
$k=\Lambda$, the inner region borders directly at the potential wall;
hence, there is no ``in-between'' that could serve as a boundary
condition. Of course, it is plausible to assume that the potential may
interpolate smoothly between the origin with zero slope and the
potential wall at $\rho=\rtc\Lambda^2+0^+$ with infinite slope. But
alternatively, the potential can also remain flat for $\rho\in
[0,\rtc\Lambda^2]$, resembling a particle-in-a-box potential.

Our ignorance about that part of the potential is unfortunately
accompanied by our inability to predict the mass of the radial
mode; but this should not come as a surprise, since the large-$N$
limit neglects the radial mode anyway.

\subsubsection{Effective potentials in $d=3$}
The investigation of the various types of potentials in $d=3$ proceeds
analogously to the $d=4$ case with almost identical results. In
particular, the symmetry-breaking potentials offer no new information:
the inner region shrinks to a point, while the potential minimum moves
to zero for $k\to 0$, and the potential wall remains at
$\rho=\rtc\Lambda$. In between, no confirmed statement can be made
within the \lN limit, since no boundary condition governs this part of
the potential. 

For symmetry-preserving potentials with $a>0$, the potential again
remains in the same form as at the cutoff for values of $\rho$ close
to the potential wall at $\rtc \Lambda$ (cf. Eq. \re{28} with
$\Lambda^2$ replaced by $\Lambda$). 

Close to the origin $\rho\to 0$, the shape of the potential is
modified; this is reflected by a mass renormalization. Employing the
same line of argument as given above in $d=4$, and using Eq. \re{A7},
we find the $d=3$ analogue of Eq. \re{29}:
\begin{equation}
1=\sqrt{\upsilon_+} \left( \frac{M}{M_\Lambda} \right)^{(a+2)/a}
\arctan \frac{1}{\sqrt{\upsilon_+}} \frac{M_\lambda}{M}. \label{33}
\end{equation}
Again, we find that there is no mass renormalization for large values
of $\upsilon_+$; corrections for small values of $\upsilon_+$ are
plotted in Fig \ref{Fig3}(b). 

\section{Conclusions}
\label{conclusions}

In the present paper, we have investigated the RG flow of particular
nonpolynomial potentials for O($N$) symmetric scalar theories using
the effective-average-action method. These Halpern-Huang potentials
arise from small relevant perturbations at the Gaussian fixed point as
tangential directions to the RG flow.
Apart from serious, unresolved
problems with the continuum limit of these potentials, we were able to
follow the flow from a given ultraviolet scale $\Lambda$ down to the
nonperturbative infrared; for this, a number of approximations have
been made which are only under limited control. In a first step, we
have neglected the influence of possible derivative couplings on the
flow of the potential.

Secondly, assuming that the anomalous dimension is only weakly
dependent on $k$ and bounded, the qualitative features of the flow
could already be guessed from the form of the flow equation: this is
because the exponential increase of the potentials essentially causes
the flow to stop for large enough field values. Therefore, the form of
the potentials was recognized as stiff under the flow; only the loose
ends of the potential near the origin or possible extrema make room
for more diversified behavior.

These considerations have been verified explicitly in the \lN limit of
the system. In this limit, the exponential increase of the potentials
is represented by a potential wall. The potential close to the wall
and the wall itself remain unchanged even in the far infrared. Those
potentials with an O($N$) symmetric ground state ($a>0$) at the cutoff
preserve this symmetry down to $k\to 0$. Our main result for such
potentials is summarized in Eqs. \re{29}, \re{30} and \re{33}, where
the particular form of the mass renormalization is stated. Contrary to
polynomial scalar interactions where the mass varies $\sim k^2$ during
the flow, the Halpern-Huang potentials exhibit corrections which are
governed by the RG distance parameter $\upsilon_+$. In particular, if
one demands that a renormalized (infrared) mass differ by several
orders of magnitude from the cutoff scale $\Lambda$, the bare
parameters of a {\em polynomial} theory at the cutoff scale have to be
fine-tuned accurately to several decimal places. By contrast, to
achieve such a separation of mass scales with a {\em nonpolynomial}
Halpern-Huang potential, an adjustment of the RG distance parameter at
the cutoff to some small value is required with much less
precision. Additionally, the smallness of this value arises naturally,
if the (unknown) perturbation at the Gaussian fixed point is tiny.
The (symmetry-preserving) Halpern-Huang potentials thus has no
problem of {\em naturalness}. Owing to the general properties of the
complete flow equation mentioned above, we believe that these
properties of the symmetry-preserving potentials in the \lN limit also
hold for finite values of $N$.

The status of the \lN limit is certainly different for Halpern-Huang
potentials which offer spontaneous symmetry breaking ($-1<a<0$). These
potentials exhibit the remarkable property that the nontrivial minimum
persists for any finite value of $k$ but vanishes in the complete
quantum theory for $k\to0$ in the \lN limit; the O($N$) symmetry is
restored and the potential becomes flat near the origin. The
coincidence between the position of the minimum and the critical value
of the field strength may finally be ascribed to the formal resemblance
between the \lN flow equation and its linearized version determining
the Halpern-Huang potentials. Since the complete flow equation
is much more complex, it appears rather improbable that this property
continues to hold for finite $N$. Therefore, whether or not
spontaneous symmetry breaking occurs in the quantum version of the
Halpern-Huang potential at finite $N$ remains an open question. The
present investigation at least observes a tendency of the system to
restore O($N$) symmetry. This is in concordance with \cite{halp98},
where a one-loop calculation for the effective potential reveals a
restoration of O($N$) symmetry for potentials with
($-1<a\lesssim-0.585$). 

In this context, a possible application of the Halpern-Huang
potentials to the Higgs sector of the standard model is still
questionable. Even if a quantum version of the potential with
spontaneous symmetry breaking exists, the naturalness of the scalar
sector alone is not sufficient to solve the hierarchy problem. This is
because the (standard) Yukawa coupling to the fermions leads to large
scalar mass renormalizations by fermion loops. Therefore, some
appropriate nonpolynomial interaction has to be chosen, also in this
sector. Nevertheless, the price to be paid would not be too high,
because not only the hierarchy problem could be circumvented without
additional degrees of freedom, but also the problem of ``triviality''
would be evaded. 


From an intuitive point of view, the fact that the form of the
potential is stable under the RG flow appears to be disappointing:
since the potential remains inherently nonpolynomial, it is impossible
to make contact with a would-be classical behavior that is determined
by only a few (polynomial) terms. The latter is usually expected at
large distances. For example, merely for very weak fields do the first
terms in a Taylor expansion of the Kummer function represent a good
approximation. For stronger fields, the application of the
Halpern-Huang potentials might therefore be limited in this sense.

From a technical viewpoint, our calculations hold for $d>2$. We have
given explicit results for $d=3$ and $d=4$, and generalizations to
higher dimensions are straightforward. The limiting case $d=2$ has to
be treated with great care for several reasons. First of all, finite
$N$ results may only be trusted if the flow of the anomalous
dimension $\eta$ is taken into account; at least in the case of
polynomial potentials, this turned out to be obligatory \cite{berg00} in
order to obtain a good picture of the Kosterlitz-Thouless
transition. Furthermore, the limit $d\to 2$ of the Halpern-Huang
potentials offers several possibilities. It has already been observed
variously in the literature (see, e.g., \cite{bonn00}), that
the Sine-Gordon as well as the Liouville potentials solve the
linearized flow equation in $d=2$. In fact, as can be easily shown
with the aid of some identities of \cite{abra}, both types of
potentials arise as limiting cases of the Halpern-Huang potentials for
$N=1$ in combination with the $\phi\to-\phi$ odd solution of the
linearized flow equation: to be precise, the Sine-Gordon potential is
recovered in the limit $d\to 2^+$ for $\lambda>2$, whereas the
Liouville potential is obtained by taking the limit $d\to 2^-$ for
$0<\lambda<2$. 

As far as the Liouville theory is concerned, further similarities to
the present results for the symmetry-preserving potentials are
visible. In \cite{reut96}, the Liouville potential has also been found
to behave stiffly under the RG flow for similar reasons as in the
present case. In particular, quantum Liouville theory appears to equal
classical Liouville theory, except for a flow of the central charge by
one unit and a modified mass parameter. These similarities confirm the
viewpoint that the Halpern-Huang potentials can be regarded as
higher-dimensional analogues of Liouville theory.

\section*{Acknowledgement}
The author wishes to thank W.~Dittrich for helpful conversations and for
carefully reading the manuscript. Useful discussions with
R.~Shaisultanov are also gratefully acknowledged. 

\section*{Appendix}

\renewcommand{\thesection}{\mbox{\Alph{section}}}
\renewcommand{\theequation}{\mbox{\Alph{section}.\arabic{equation}}}
\setcounter{section}{0}
\setcounter{equation}{0}

\section{Integrals for the \lN flow equation}
In this appendix, we present some details about the function
$I(d,t;\du)$ appearing in the solution \re{16} to the flow equation
\re{15}; this function is defined as
\begin{equation}
I(d,t;\du):= \E^{-(d-2)t} \int\limits_0^{\exp(-2t)} dw\,
\frac{w^{-d/2}}{1+\E^{2t} \, \du\, w}. \label{A1}
\end{equation}
Substituting $w=\exp[-2(T+t)]$, we arrive at the form
\begin{equation}
I(d,t;\du)=2 \int\limits_0^{-t} dT\, \frac{\E^{(d-2)T}}{1+\du \,
  \E^{-2T}}, \label{A2}
\end{equation}
where $t=\ln k/\Lambda$ is always nonpositive:  $t\in]-\infty,
0]$. Separating the zeroth-order term of a Taylor expansion of the
integrand, we find the convenient representation
\begin{equation}
I(d,t;\du)=i_0(d,t)-\du\, J(d,t;\du), \label{A3}
\end{equation}
with the auxiliary functions $i_0(d,t)$ and $J(d,t;\du)$ defined by
\begin{eqnarray}
i_0(d,t)&:=& \frac{2}{d-2} \bigl(\E^{-(d-2)t} -1\bigr),\label{A4}\\
J(d,t;\du)&:=&2 \int\limits_0^{-t} dT\, \frac{\E^{(d-4)T}}{1+\du\,
  \E^{-2T}}. \label{A5}
\end{eqnarray}
Note that $i_0,J\geq 0$ for $t\leq 0$ and $\du>-1$. The explicit form
of $J$ depends on the spacetime dimension. For $d=4$, the integral can
easily be evaluated by standard means, yielding
\begin{equation}
J(4,t;\du)=\ln \frac{\E^{-2t}+\du}{1+\du}. \label{A6}
\end{equation}
In $d=3$, we take care of the possibility of a nontrivial minimum
(spontaneous symmetry breaking) and find to the right of a possible
minimum
\begin{equation}
J(3,t;\du>0)=-\frac{2}{\sqrt{\du}} \left( \arctan \frac{1}{\sqrt{\du}}
  - \arctan \frac{\E^{-t}}{\sqrt{\du}} \right). \label{A7}
\end{equation}
In the ``inner'' region to the left of a possible minimum, we obtain
\begin{equation}
J(3,t;\du<0)=\frac{2}{\sqrt{-\du}} \left( \text{Artanh}
  \frac{1}{\sqrt{-\du}}  - \text{Artanh} \frac{\E^{-t}}{\sqrt{-\du}}
  \right), \label{A8} 
\end{equation}
where $\du>-1$ for reasons of consistency.

\end{document}